# YeastMed: an XML-Based System for Biological Data Integration of Yeast

Abdelaali Briache[1], Kamar Marrakchi[1], Amine Kerzazi[2], Ismael-Navas-Delgado[2], José F. Aldana-Montes[2], Badr D. Rossi Hassani[1] and Khalid Lairini[1]

[1]Department of Biology, Faculty of Sciences and Techniques, BP: 416 - University Abdelmalek Essaâdi, Tangier. Morocco.
[2]Department of Computer Languages and Computing Science, Higher Technical School of Computer Science Engineering University of Málaga, 29071, Malaga, Spain,
{a_briache, mkamar, kerzazi, ismael, jfam}@lcc.uma.es
badrrossi@gmail.com, Khalidlairini@yahoo.com

**Abstract.** A key goal of bioinformatics is to create database systems and software platforms capable of storing and analysing large sets of biological data. Hundreds of biological databases are now available and provide access to huge amount of biological data. SGD, Yeastract, CYGD-MIPS, BioGrid and PhosphoGrid are five of the most visited databases by the yeast community. These sources provide complementary data on biological entities. Biologists are brought systematically to query these data sources in order to analyse the results of their experiments. Because of the heterogeneity of these sources, querying them separately and then manually combining the returned result is a complex and laborious task. To provide transparent and simultaneous access to these sources, we have developed a mediator-based system called *YeastMed*. In this paper, we present *YeastMed* focusing on its architecture.

**Keywords:** Semantic Web, YeastMed, SB-KOM, Yeast, Web Services, Data Integration.

## 1 Introduction

The most well-known and commercially significant yeasts are the related species and strains of Saccharomyces cerevisiae. Nowadays, the word yeast is widely given to the species Saccharomyces cerevisiae because of the place it takes in Life Science. Saccharomyces cerevisiae is now recognized as a model system representing a simple eukaryote whose genome can be easily manipulated. Life Science is generating large amounts of data concerning Saccharomyces cerevisiae that have been stored in multiple databases. Biologists are brought systematically to query these sources in order to analyse the results of their experiments. They must perform at least the following tasks during query formulation and execution: (i) identify appropriate sources and their location, (ii) identify the focus of each source, (iii) query each convenient source independently using its specific access method and query language, (iv) navigate through the sources to obtain complementary data, and (vi) manually merge the results obtained from different sources. This places a burden on biologists, most of whom are not bioinformatics experts, and limits the use that can be made of the available information.

The challenges of modern bioinformatics research is not only storing data in repositories, but also processing and integrating them. Multiple solutions to biological data integration have been developed. Researchers have come up with some approaches that integrate diverse biological data sources. There are mainly two integration approaches being used in addressing the issue of the interoperability between biological databases: data warehousing approach [1] and mediator-based approach [3].

The data warehousing approach is adopted by numerous integration systems like GUS [2] and DiscoveryLink [4]. This approach uses a data warehouse repository that provides a single access point to a collection of data, obtained from a set of distributed, heterogeneous sources. Data from the remote heterogeneous databases are copied on a local server and the user will use a unique interface within the system to allow multi-database queries to be issued to this single interface.

A mediator does not store any data, but it provides a virtual view of the integrated sources. The mediator approach basically translates the user query, into queries that are understood by the integrated sources. It maps the relationship between source descriptions and the mediator and thus allows queries on the mediator to be translated to queries on the data source.

The mediator-based approach has several strengths compared to data warehouse. It does not have the updating problem as the query goes directly to the original source. Mediators can be seen as a cheaper and more effective approach since they use schema or view integration, rather than having to have huge storage capacity to store copied data from all the data sources involved.

This paper presents a mediator-based system called *YeastMed* that aims to provide transparent access to disparate biological databases of yeast. It provides a unique interface between the user who submits a query, and a set of five data sources accessible via web protocols. *YeastMed* relies on SB-KOM [10] to perform the query transformation needed to reach the integrated data sources. These sources are the most visited databases by the yeast community: *SGD* [5], *Yeastract* [6], *CYGD-MIPS* [7], *BioGrid* [8] and *PhosphoGrid* [9]. They provide complementary data on biological entities (cellular interaction, metabolic pathways, transcription factors, annotation data...). With *YeastMed*, we aim to help biologists to understand and explain the biological processes of interest by using an integrative system.

This paper is organized as follows: an overview on some biological data integration systems is given in the next section. In Section 3, a general overview of the system and the resources used in *YeastMed* are described. In Section 4, the integration process along with some explanatory schemas is presented. The data mapping rules that have been defined for instances reconciliation during the integration process are also presented. A detailed use case is then given in Section 5 to describe how *YeastMed* proceeds when a user query is submitted. Section 6 concludes the paper.

## 2   Related work

The evolution of biological data integration has given birth to several systems that regroup and integrate a set of heterogeneous databases. But in the context of yeast systems, the area has not yet had a remarkable evolution. We can cite Cell Cycle Database [12] and YeastHub [11].

- Cell Cycle Database [12] is an integrated data warehouse for systems biology modeling and cell cycle analysis based on yeast and mammalian organisms. The system integrates information about genes and proteins involved in the cell cycle process. It stores complete models of interaction networks and allows the mathematical simulation over time of quantitative behaviour of each component. The database integration system consists of a series of programs used to retrieve the data from several different external databases, transform and load them into the warehouse data model; it also consists of a series of links with external resources which allows a wider exploration of available information about cell cycle components.
- YeastHub is a prototype application in which a data warehouse has been constructed in order to store and query different types of yeast genome data provided by different resources in different formats including the tabular and RDF formats. Once the data are loaded into the data warehouse, RDF-based queries can be formulated to retrieve and query the data in an integrated fashion. YeastHub is implemented using Sesame 1.1 and uses Tomcat as the web server. The web interface is written using Java servlets. The tabular-to-RDF conversion is written using Java. To access and query the repository programmatically, it uses Sesame's Sail API that is Java-based. It uses MySQL as the database server (version 3.23.58) to store information about the correspondences between the resource properties and the query form fields.

These two systems present some limitations: the first one is that they store the extracted data locally in a data warehouse or database which render the updating process a tedious task. YeastHub presents another problem where Sesame does not have a way to identify the source of the triples (statements) once they are loaded into the repository.

## 3   YeastMed overview

*YeastMed* is a mediator-based system that consists of several components contributing to the data integration process in different ways. In this section we settle for an insight into its architecture and also the set of the data sources it integrates, while in the next section we talk in detail about its components and the role of each of them in the system.

### 3.1 General Architecture

The general architecture of the *YeastMed* system is shown in Figure 1. It consists of a set of components that have been implemented independently and play different roles. The access point to the system is a web interface that incorporates two search forms: a quick search form and an advanced search form. Scientists use the former to quickly submit their requests based on some keywords (Gene or Protein names, GO terms or any other words that can appear in the search fields of the interrogated data sources). The latter is an ontology-based form. Using it, biologists can express their requests in terms of the *YeastMed* Ontology.

*YeastMed* relies on SB-KOM [10] (an ontology-based mediator developed by Khaos group of the University of Malaga) to perform query transformation at execution time. Once the user submits a request from the web interface, *YeastMed* generates a conjunctive query. SB-KOM decomposes this query into suitable sub-queries to individual sources based on a set of mapping rules. These sub-queries are expressed in XQuery.

*YeastMed* have a set of web services (Data services for us): one for each integrated source. These components receive XQueries from SB-KOM and return XML documents. The role of the web services is to allow *YeastMed* to use wrapper functionalities to find and extract solicited information from data sources through their web pages or FTP mechanisms. Answers, materialized by XML documents, to XQueries are sent to the mediator which combines them into a YeastMed ontology instance expressed in RDF. The final result is provided to the user in different formats (RDF, XML or HTML).

Data sources are also an important component in the *YeastMed* architecture because they are the providers of the biological information.

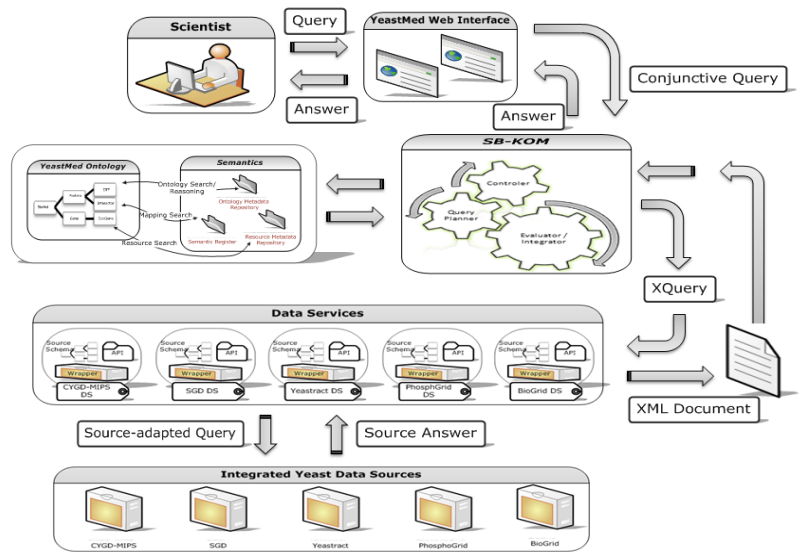

**Fig. 1.** General Architecture of *YeastMed* System.

### 3.2 Integrated Data Sources

In its current version, *YeastMed* integrates five Yeast databases. They have been selected for having the most appropriate properties for studying Saccharomyces cerevisiae. These sources provide complementary data concerning genome, proteome, metabolome and reactome of Saccharomyces cerevisiae:

- SGD Database [5] (http://www.yeastgenome.org/): It contains the sequences of yeast genes and proteins, descriptions and classifications of their biological roles, molecular functions, subcellular localisations, links to literature information and tools for analysis and comparison of sequences.
- YEASTRACT Database [6] (http://www.yeastract.com): It is a repository of regulatory associations between transcription factors and target genes, based on experimental evidence which was spread throughout bibliographic references. Each regulation has been annotated manually, after examination of the relevant references. The database also contains the description of specific DNA binding sites for a sub-group of transcription factors.
- MIPS-CYGD [7] (http://mips.helmholtz-muenchen.de/genre/proj/yeast/): aims in general to present information on the molecular structure and functional network of Saccharomyces cerevisiae. In addition, the data of various projects on related yeasts are also used for comparative analysis.
- BioGRID [8] (http://thebiogrid.org/): It is an online interaction repository with data compiled through comprehensive curation efforts. All interaction data are freely provided through the search index and available via download in a wide variety of standardized formats.
- PhosphoGRID [9] (www.phosphogrid.org): records the positions of specific phosphorylated residues on gene products. Where available for specific sites, PhosphoGRID has also noted the relevant protein kinases and/or phosphatases, the specific condition(s) under which phosphorylation occurs, and the effect(s) that phosphorylation has on protein function.

## 4 Components of Biological data integration in YeastMed

*YeastMed* has a set of modules that depend heavily on XML technology to integrate syntactically and semantically biological data. In what follows, we give detailed information on these components.

### 4.1 Source Schemas

The knowledge modeling of the application domain of *YeastMed* constitutes the corner stone towards an efficient integration. To that end, a detailed study of the sources has been carried out with the goal of establishing a standard terminology to describe the data. Each data source has been modeled by an exported XML schema (see an example in Figure 2). An exported schema refers to translated source schema in the *YeastMed* Ontology. These schemas are considered as models describing data and their organization in data sources and define a structure under which results will be returned by web services.

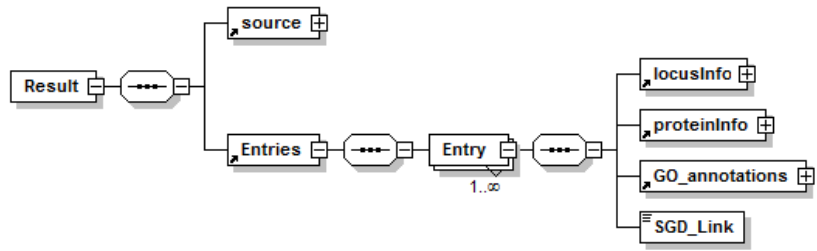

**Fig. 2:** a fragment of the Yeastract Schema.

### 4.2 Data Services

*YeastMed* uses a set of web services (called in our case Data Services) to access data sources. We have developed one Data Service for each integrated yeast source. These components hide technical and data model details of the data source from the mediator. They receive XQueries from SB-KOM and return XML documents in addition to other metadata. The role of *YeastMed* Data Services is double:

- Allowing *YeastMed* to use the wrapper functionalities to find and extract solicited information from data sources using HTML protocols or FTP mechanisms. This means providing the ability to solve XQueries and return answers in XML format.
- Exporting semantic information about data schemas and data provenance. This allows mainly *YeastMed* to keep track of the returned information when combining them and which source is being interrogated.

It is commonly known that a wrapper is an interface to a data source that translates data into the common data model used by mediators [16]. Since the goal of YeastMed is to integrate databases accessible via Web protocols, it is completely normal that a wrapper is considered as the most important component of the architecture of *YeastMed* Data Services. It is an interface that receives XQueries generated by SB-KOM, accesses a specific data source, extracts data and translates them into the common data model used by SB-KOM, i.e. XML (see Figure 3).

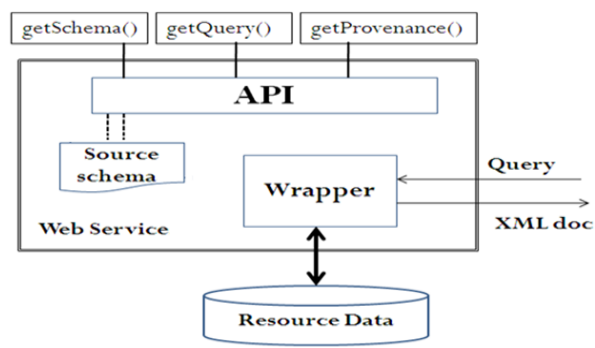

**Fig. 3:** Architecture of the web services in *YeastMed* System.

In addition to the wrapper's query service, the web services encapsulate an Application Programming Interface (API). It is the access point for SB-KOM to the functionality of the web service. This API publishes three methods: *Query(Q)* that passes to the wrapper the XQuery *Q* and returns its answer in an XML format. The XML structure of this answer must satisfy the constraints of the source schema. The other two methods, *getSchema()* and *getProvenance(),* provide access to the metadata that the web service stores. The former returns the XML data schema and the latter provides information on the underlying data source. In order to use these methods correctly, SB-KOM finds all the necessary information about them in a WSDL (Web Service Description Language) document.

The web services have been implemented in Java. They receive XQueries from SB-KOM via the *getQuery()* method of the API which passes it to the wrapper. This is materialised by a set of java classes that define several methods. The incoming XQuery is analysed to identify precisely what information is solicited from the underlying data source. The wrapper then generates a source-adapted query following the query capabilities of the source. Then it establishes a connection to the data source via HTML or FTP protocols and submits the query to it. A set of methods are defined to extract data from source answers and organize them as an instance of the XML source schema before sending it to the SB-KOM in the form of an XML document.

YeastMed is able to reflect data provenance by calling the method *getProvenance()* which returns information about a source or through the XML document returned by data services: it contains by default a description of the interrogated data source. Thus, instances with the integrated data can be annotated with the data provenance of each piece of information. In this way, the user interface could show users the provenance of each part of the results.

### 4.3 YeastMed Ontology

Since we aim to help scientists to get information from multiple sources by providing a single access point, we have equipped *YeastMed* with a domain ontology. This ontology is the product of the reconciliation of the different data source schemas into a single, coherent schema. It is a knowledge model that captures biological and bioinformatics knowledge in a hierarchical conceptual framework constrained by parent-child relationships: A child (see Figure 4) is a subset of a parent's elements; each child inherits all of its parent's properties but has more specialized properties of its own.

In this ontology, we have created a class hierarchy which is a classification of all the biological entities manipulated by the system. In addition, the ontology defines two types of properties which are used to convey additional semantic information about the concepts of the ontology. While the first one is defined by a set of object properties that model the relationships that can hold between two individuals belonging to one or two different classes of the ontology, the second type are data properties: these are relationships linking an individual to a literal data.

The primary purpose of *YeastMed* Ontology is to support the user queries. Queries are phrased in terms of the ontology terms and *YeastMed* converts these to XQuery requests to the appropriate sources. This process is based on the mappings rules.

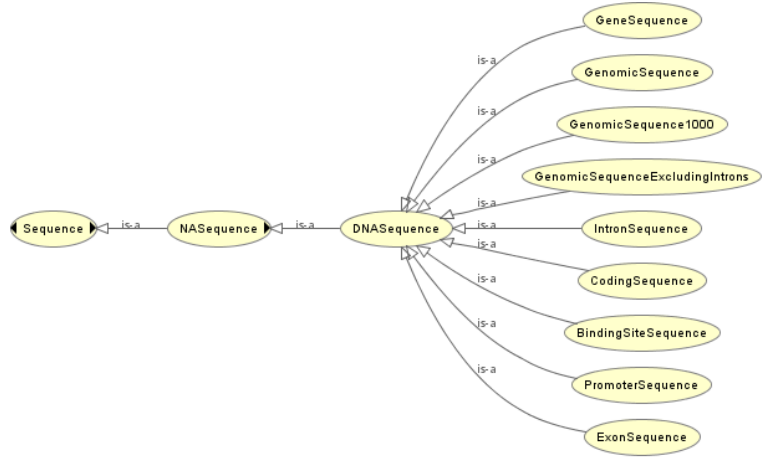

**Fig. 4:** a fragment of the *YeastMed* Ontology. It shows all the ancestors and the children of the DNASequence Concept.

### 4.4 Mappings

Having a domain ontology facilitates the formulation of queries to the system. The users simply pose queries in terms of the ontology rather than directly in terms of the Source Schemas. Although this is very practical and effective in terms of the system transparency to the user, it brings the problem of mapping the query in the mediated schema to one or more queries in the schemas of the data sources. In *YeastMed*, this problem is solved using the functionality of SB-KOM. So in addition to modeling the ontology and the sources, we needed to establish associations between the concepts in the ontology and the appropriate elements representing the information in the sources. These associations are materialized in *YeastMed* by the mapping rules.

SB-KOM is designed to decompose queries based on GAV approach-based mappings [17]. That means each concept (also property in our case) in the ontology is a view defined in terms of the source schemas' elements. This view specifies how to obtain instances of the mediated schema elements from sources. In this context, the mapping rules we have used are defined as pairs (P,Q). P is one or a couple of path expressions on a source schema expressed in XPath, and *Q* a conjunctive query expressed in terms of the Ontology terms. Three kinds of mappings have been defined:

- Class Mapping: it maps ontology classes to source schemas. It has the following form:
  ```
  XPath-Element-Location,Ontology-Class-Name,
  correspondence-index
  ```
  Where *XPath-Element-Location* is the location of an element in the source schema, expressed in XPath; *Ontology-Class-Name* is the name of the corresponding class in the Ontology and *correspondence-index* is an integer value that informs on the correctness of the mapping instance. In *YeastMed*, this index is always 100 since all the mappings are done manually and not

automatically. An example which maps the *Protein* class to the SGD schema is as follows:
```
Result/Entries/Entry/Protein, Protein,100
```

- Datatype Property Mapping: it maps ontology datatype properties to source schemas. It has the following form:
```
XPath-Domain-Location; XPath-value-Location,
Ontology-Domain-Name; Property-Name,
correspondence-index
```
*XPath-Domain-Location* is the Path to the element in the source schema which is mapped to the domain of the datatype property; *XPath-value-Location* is the Path to the element where the property takes the value of its range and *Ontology-Domain-Name* and *Property-Name* are respectively the domain and the name of the property.

- Object Property Mapping: it maps ontology object properties to source schemas. It has the following form:
```
XPath-Domain-Location; XPath-Range-Location,
Ontology-Domain-Name; Ontology-Range-Name;
Property-Name, correspondence-index
```
*XPath-Range-Location* is the Path to the element in the source schema which is mapped to the range of the object property. *Ontology-Range-Name* is the range name of the property.

### 4.5 SB-KOM

*YeastMed* relies on SB-KOM, which is based on KOMF [13], to perform query transformations at execution time. KOMF is a generic infrastructure to register and manage ontologies, their relationships and also information relating to the resources. This infrastructure is based on a resource directory, called Semantic Directory [14], with information about web resource semantics. KOMF has been successfully instantiated in the context of molecular biology for integrating biological data sources which are accessible via internet pages. SB-KOM mediator is composed of three main components: the Controller, the Query planner and the Evaluator/Integrator.

#### 4.5.1 Controller

The controller component receives requests coming from *YeastMed* web interface and evaluates them to obtain a result for the requests. The controller creates different threads for different user requests, and assumes the role of the middleware between the mediator components.

Queries are expressed as conjunctive predicates [15], with three main types of predicate: classes in terms of *YeastMed* ontology which is registered in the Semantic Directory, datatype properties that link individuals to data values, and object properties that link individuals to individuals. The results of these queries are instances of the *YeastMed* ontology which the query was expressed in.

### 4.5.2 Query Planner

This component is by far one of the most fundamental pillars in elaborating one or several query plans to solve the query from different data sources. Plans generated by this component specify the data sources from which the information can be retrieved and in which order they must be accessed. The evaluation of these queries depends on the query plans themselves.

According to the query (a Conjunctive Query), there will be different types of mapping in the semantic directory. Classes will be connected to the XPath of one or several XML Schema resource elements. On the other hand, datatype properties will be connected to those two expressions: the first one corresponds to the class and the second to the property. The object properties will be related to the active XPath classes in the property.

### 4.5.3 Evaluator/Integrator

This component analyses the query plan (QP), and performs the corresponding calls to the data services involved in the sub-queries (SQ1, ...,SQn) of the query plan. To answer *YeastMed* query, this component first executes the data services in the order specified by the query plan. Then, it obtains the instances from the data service results. These instances are not interconnected because they have been produced by different data services. In order to retrieve a set of interrelated instances we need to establish relationships between them. This can be achieved by the object properties defined in the ontology that are used as relationships between services in the query plan. Finally, these interrelated instances are filtered in order to eliminate the information not required.

## 5 Use Case

In this section, we show how a user query is solved by *YeastMed*, and how its different components take part in this process. Let us take the case of a biologist who is using *YeastMed* to find information about two kinds of proteins. The first one is represented by DNA Topoisomerase III, and the second one is indicated by some transcription factors regulating the expression of the first kind. The biologist is interested in the phosphorylation sites that are found in the sequences of the transcription factors of DNA Topoisomerase III, especially the one (or ones if they exist) whose gene is located on the Chromosome XVI. In addition, the biologist also aims to get all the literature on DNA Topoisomerase III. As stated previously, *YeastMed* provides a web interface that allows biologists to express their queries in terms of the ontology (see Figure 5).

**Fig. 5:** A part of the ontology-based search form.

The fragment of semantics that is implied directly in the formulating process of that query is shown in Figure 6. From this fragment, a conjunctive query is generated automatically:

```
Ans(BR,Ph):= Protein(P), hasDescription(P,"DNA
Topoisomerase III"),BibRef(BR),hasBibRef(P,BR),
hasSystematicName(P,SN),regulatedBy(P,TF),hasName(T
F,Nt),TranscriptionFactor(TF),Chromosome(C),hasName
(C,"XVI"), BelongsTo(TF,C),PhosphoSite(Ph),
hasPhosphoSite(TF,Ph);
```

This conjunctive query includes as predicates five ontology classes (*Protein*, *BibRef*, *TranscriptionFactor*, *Chromosome* and *PhosphoSite*), three datatype properties (*hasDescription*, *hasSystematicName* and *hasName*) and four object properties (*hasBibRef*, *regulatedBy*, *belongsTo* and *hasPhosphoSite*). This query will return instances of *PhosphoSite* and *BibRef* that satisfy its constraints.

As a subsequent step, the conjunctive query will be sent to SB-KOM, received by the controller which will pass it to the Query Planner. This component has an algorithm that, based on the query predicates and the mappings of the semantic directories, will generate a set of sub-queries and also a plan to execute them. The predicates of the conjunctive query are divided into two sets: a set that contains predicates with a single argument and another that contains predicates with more than one argument. The predicates from the two sets having common arguments are then grouped together into groups represented by the combination of two or more predicates. The groups that are not represented in the Semantic Directory mappings are discarded (see Figure 6 for the mappings implied in this process). The remainder is added to the first set allowing a group to be present only once. Table 1 lists all resulting groups.

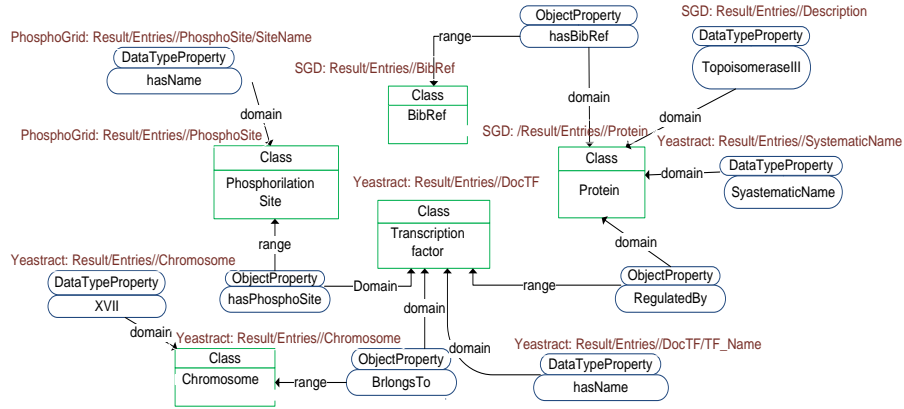

**Fig. 6:** The fragment of the Ontology invoked to formulate the query example. Classes are shown in green and Properties in blue. The mappings between the ontology and the source schemas are present above the ontology element (in red).

**Table 1.** The groups used to form the Plan Tree. For each group the Mapping Source is given.

| Group | Query | Mapping source |
|---|---|---|
| G1 | Protein(P), hasBibRef(P,BR) | SGD |
| G2 | Protein(P),hasDescription(P,"DNA Topoisomerase III") | SGD |
| G3 | Protein(P), hasSystematicName(P, SN) | Yeastract |
| G4 | Protein(P), RegulatedBy(P, TF) | Yeastract |
| G5 | TranscriptionFactor(TF), hasName(TF, Nt) | Yeastract |
| G6 | TranscriptionFactor(TF), belongsTo(TF,C) | Yeastract |
| G7 | TranscriptionFactor(TF), hasPhosphorylationSite(TF, Ph) | PhosphoGrid |
| G8 | Chromosom(C), hasName(C,"XVI") | Yeastract |
| G9 | regulatedBy(P,TF) | Yeastract |
| G10 | hasBibRef(P,BR) | SGD |
| G11 | belongsTo(TF,C) | Yeastract |
| G12 | hasPhosphoSite(TF,Ph) | PhosphoGrid |
| G13 | Protein(P) | SGD; Yeastract; PhosphoGrid |
| G14 | TranscriptionFactor(TF) | Yeastract; PhosphoGrid |
| G15 | BibRef(BR) | SGD |
| G16 | Chromosome(C) | Yeastract |
| G17 | PhosphoSite(Ph) | PhosphoGrid |

From this set, the planner will try to construct potential trees of the execution order. It selects groups having variables instantiated in order to set a root for a tree. The order of the plan execution depends on the instantiated variables: the group containing an instanced variable is executed first, then the groups that are related to those variables, and so on until all the groups are executed. In our case, G2 and G3 are selected. G8 cannot serve as a root, because there is no other group that depends on its instantiated variable which keeps the other groups without execution. This is not the case for G2 which serves as a root for the tree shown in Figure 7. It is the first to be executed. This returns the protein that has as description "DNA Topoisomerase III". Then G9 and G10 are executed in parallel because they depend on the instantiated variable of G2. From these simultaneous executions, the algorithm will determine all the objects that are related to *Protein* by means of the relationships *regulatedBy* and *hasBibRef*. Once those objects are obtained, it will check whether they satisfy G14 and G15: that means checking if the objects obtained from G9 and G10 are respectively of the type *TranscriptionFactor* and *BibRef*. Based on the result of G9, groups G11 and G12 are executed but not simultaneously. SB-KOM has a plan optimization module that might change the order of the initial plan execution as is the case here: Since G8 has a variable instantiated (value "XVII") and is related to G14 via G11, this one is executed before G12, and the result is used by this group to be executed. The arcs of the planning trees generated by the planer represent object properties, while the nodes are ontology concepts or instances of these. Each node and arc contains all the necessary information for the Evaluator /Integrator to execute sub-queries. That is: the XQuery (elaborated from the mapping) corresponding to the sub-query of the node or the arc, the names and the URLs of the Data Service of interest. An example is shown in Figure 8.

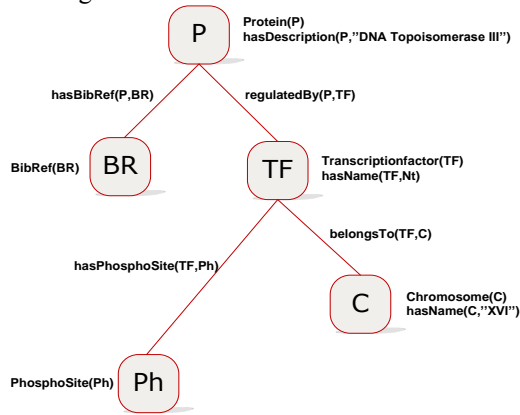

**Fig. 7:** The plan tree generated from the conjunctive query

```
Ontology: YeastMedOntology, http://khaos.uma.es:8088/ontologies/YeastMedOntology
Resource: SGD, http://khaos.uma.es:8088/services/SGDService
Ontology terms: Protein; hasDescription,
Resource elements: /Result/Entries/Entry/Protein; /Result/Entries/Entry/Protein/Description,
Xquery:
        for $d in /Result/Entries/Entry/Protein
        where $d/Description eq "DNA Topoisomerase III"
        return $d
```

**Fig. 8:** The information presented by the node P in the plan tree sketched in this section.

The *YeastMed* data services are executed by the Evaluator/Integrator following the plan, after optimization, generated by the Planner. In our case, SGD Data Service receives the first sub-query, because the object property *hasDescription* is mapped to the SGD Schema. TOP3 is returned as an answer of this sub-query and then is used by the sub-query *RegulatedBy* to find instances of *TranscriptionFactor*. The Yeastract Data Service is invoked this time because the property is mapped to the Yeastract Schema. Three instances of the type *TranscriptionFactor* are returned: Fhl1p, Hsf1p and Swi4p. For each of these instances, the Yeastract Data Service is called again. It receives this time the sub-query represented by the property *belongsTo* that contains the two arguments instantiated: the first one is one of the three instances returned by the previous query, and the second argument is instantiated by the name of the chromosome XVI. This sub-query checks whether the Transcription factor has its coding gene on the chromosome XVI. Only the instance Fhl1p is maintained. Finally the sub-query hasPhosphoSite is executed on the PhosphoGrid Data Service that returns all the *PhosphoSite* instances of the Transcription Factor Fhl1p. At each execution, the Evaluator/Integrator receives results in XML format from the target Data Services. These results are instances of the XML schemas of the underlying sources. Based on the mapping between the elements of the source schemas and the elements of the ontology, these XML schema instances are translated into ontology instances which are not interconnected because they have been produced by different data services. To associate them, the Evaluator/Integrator uses just the instances of the domain and range classes of the object properties. The final result is an ontology instance that includes all the data extracted from the interrogated data sources. That is all the instances of the concepts BibRef of the protein TOP3 and all the PhosphoSite objects of the Transcription Factor Fhl1p.

## 6   Conclusions

We have described *YeastMed*: an XML and mediator-based system that Integrates five Yeast databases which have the most appropriate properties for studying Saccharomyces cerevisiae.

Data services play an important role in the integration process of this system, where they are considered as an interface which receives queries, accesses to a data source, extracts data and translates them into a common data model used by SB-KOM. In YeastMed, Data Services extract data mainly from flat files because most of the integrated data sources are accessible via ftp mechanisms and provide data in tabular or XML format. This reduces the costs of the maintainability of the system because flat files structures are not frequently target to changes.

In our system, the schema integrator is an ontology and the results are ontology instances. The use of the ontology and instances enables reasoning to be later included at different levels. YeastMed could be equipped with a reasoner that would infer new relationships between the instances of the ontology when solving a user query. This will enable YeastMed to discover new knowledge for the query answers. The final result is an ontology instance that includes all the data extracted from the integrated data sources.

**Acknowledgements**

This work has been funded by the project PCI-EUROMED A/016116/08, P07-TIC-02978 (Junta de Andalucía), and TIN2008-04844 (Spanish Ministry of Science and Innovation).